\pdfoutput=1

\documentclass[aps,prb,reprint,showpacs,superscriptaddress,byrevtex]{revtex4-1}

\usepackage{graphicx}
\usepackage{graphics}
\usepackage{amsmath}
\usepackage{amssymb}
\usepackage{amsfonts}
\usepackage{dcolumn}
\usepackage{dsfont}
\usepackage{latexsym}
\usepackage{rotating}
\usepackage{color}
\usepackage{latexsym}
\usepackage{bbm}
\usepackage{subfigure}
\usepackage{float}
\usepackage{epsfig}
\usepackage{epsf}
\usepackage{psfrag}
\usepackage{bm}
\usepackage{amsthm}
\usepackage{eucal}
\usepackage{mathrsfs}
\usepackage{url}
\usepackage{braket}
\usepackage{natbib}
\usepackage{multirow}
\usepackage{color} 

\usepackage{hyperref}
\hypersetup{
colorlinks=true,final=true,
        linkcolor=blue,
        citecolor=blue,
        filecolor=blue,
        urlcolor=blue,
}

\begin{document}
\title{Orbitally driven spin reorientation in Mn doped YBaCuFeO$_{5}$}

\author{Mukesh Sharma}
\author{T. Maitra}
\email{tulika.maitra@ph.iitr.ac.in}
\affiliation{Department of Physics, Indian Institute of Technology Roorkee, Roorkee-247667, Uttarakhand,India}

\begin{abstract}
Oxygen-deficient layered perovskite YBaCuFeO$_{5}$ (YBCFO) is one rare type-II multiferroic material where ferroelectricity, driven by incommensurate spiral magnetic order, is believed to be achievable up to temperatures higher than room temperature. A cycloidal spiral rather than helical spiral order is essential ingredient for the existence of ferroelectricity in this material. Motivated by a recent experimental work on Mn-doped YBCFO where the spiral plane is observed to cant more towards the crystallographic $c$ axis upon Mn doping at Fe sites compared to that in the parent compound, we performed a detailed theoretical investigation using the density functional theory calculations to understand the mechanism behind such spin reorientation. Our total energy calculations, within GGA+U+SO approximation, reveal that Fe/Cu spin moments indeed align more towards $c$ axis in the Mn-doped compounds than they were in parent compound YBCFO. Largest exchange interaction (Cu-Cu in the $ab$-plane) is observed to decrease systematically with Mn doping concentration reflecting the lowering of transition temperature seen in experiment. Further, the inter-bilayer Cu-Mn exchange interaction becomes ferromagnetic in the doped compound whereas the corresponding Cu-Fe exchange was antiferromagnetic in the parent compound giving rise to frustration in the commensurate magnetic order. Most importantly, our electronic structure calculations reveal that in the doped compounds because of hybridization with the d$_{z^2}$ orbital of Mn, the highest occupied Cu orbital becomes d$_{z^2}$ as opposed to the parent compound where the highest occupied Cu orbital is d$_{x^2-y^2}$. Therefore, we believe that the occupancy of out of plane oriented Cu d$_{z^2}$ orbital in place of planar d$_{x^2-y^2}$ orbital along with the frustrating exchange interaction drive the spins to align along $c$ in doped compound. 

\end{abstract}
\maketitle
\section{Introduction}

Multiferroic materials with high transition temperature as well as strong coupling between magnetization and polarization are highly sought after because of their tremendous potential in device applications such as in spintronics, storage devices, sensors etc.\cite{Yoshinori_Tokura2006, W_Eerenstein2006}. Type-I multiferroics, where magnetization and polarization have different origins, are found to have high transition temperatures but due to weak coupling between magnetization and polarization these materials are less useful. In contrast, the type-II multiferroics where the ferroelectricity is driven by the magnetic ordering, the coupling between magnetization and polarization is strong and hence are more useful for device applications\cite{Buurma2016, Khomskii2006}. Unfortunately, most type-II multiferroics have much lower transition temperatures compared to room temperature\cite{Khomskii2009}.    

Cupric oxide (CuO), and bilayer perovskite, YBaCuFeO$_5$ (YBCFO), are the two so far discovered type-II multiferroics, which show ferroelectric ordering at higher temperatures ($>$ 200 K). In CuO, a spiral magnetic order drives the ferroelectricity which survives up to higher temperatures but in a narrow temperature window (213 K - 230 K)\cite{T_kimura2008}. YBCFO\cite{L_Er-Rakho1988, B_Kundys2009, Yuji_Kawamura2010}, on the other hand, has spiral (incommensurate) magnetic ordering up to a much higher temperature and in a wide range (T $<$ 230 K). YBCFO displays two magnetic transitions, paramagnetic to commensurate (CM) antiferromagnetic (T$_{N1}\sim$ 440 K) and CM to incommensurate (ICM) (T$_{N2}\sim$ 230 K) spiral magnetic ordering, respectively\cite{B_Kundys2009}. This ICM or spiral magnetic order is believed to be accompanied by ferroelectricity as some earlier experimental work reported\cite{B_Kundys2009, Yuji_Kawamura2010, morin2015}. 

Of late, there has been a debate on the nature of spiral magnetic state (whether helical or cycloidal) and existence of ferroelectricity in this material\cite{Lai2017, Dey2018}. Cycloidal spiral state (where the spin moments are directed out of $ab$-plane) is necessary to have finite polarization in this system\cite{Dey2018}. In non geometrically frustrated system which is the case in YBCFO, competing nearest (J$_{NN}$) and next nearest neighbour (J$_{NNN}$) exchange interaction along $c$ direction can stabilize a cycloidal spiral state. However, it has been found from density functional theory (DFT) calculations that neither (J$_{NNN}$) nor the Dzyaloshinskii-Moriya interaction (DMI) is strong enough to give rise to such a state \cite{morin2015, Dey2018}. An alternative mechanism involving Fe/Cu chemical disorder giving rise to randomly distributed Fe-Fe antiferromagnetic bonds was subsequently proposed for causing the desired frustration in the commensurate magnetic structure and stabilizing the spiral order thereby\cite{morin2015, Scaramucci2018}. Later on, experimentally it was shown that by introducing Fe/Cu chemical disorder during the preparation of sample indeed gives rise to stable spiral order up to a much higher temperature\cite{morin2016}. A combination of both chemical disorder and substitution at the Ba sites by other rare-earth ions has been shown recently to raise the spiral ordering temperature to way beyond room temperature\cite{shang2018, Lal2017}. However, these authors did not report any observation of ferroelectricity in their samples. 

\begin{figure}[ht!]
    \centering
    \includegraphics[width = 8 cm]{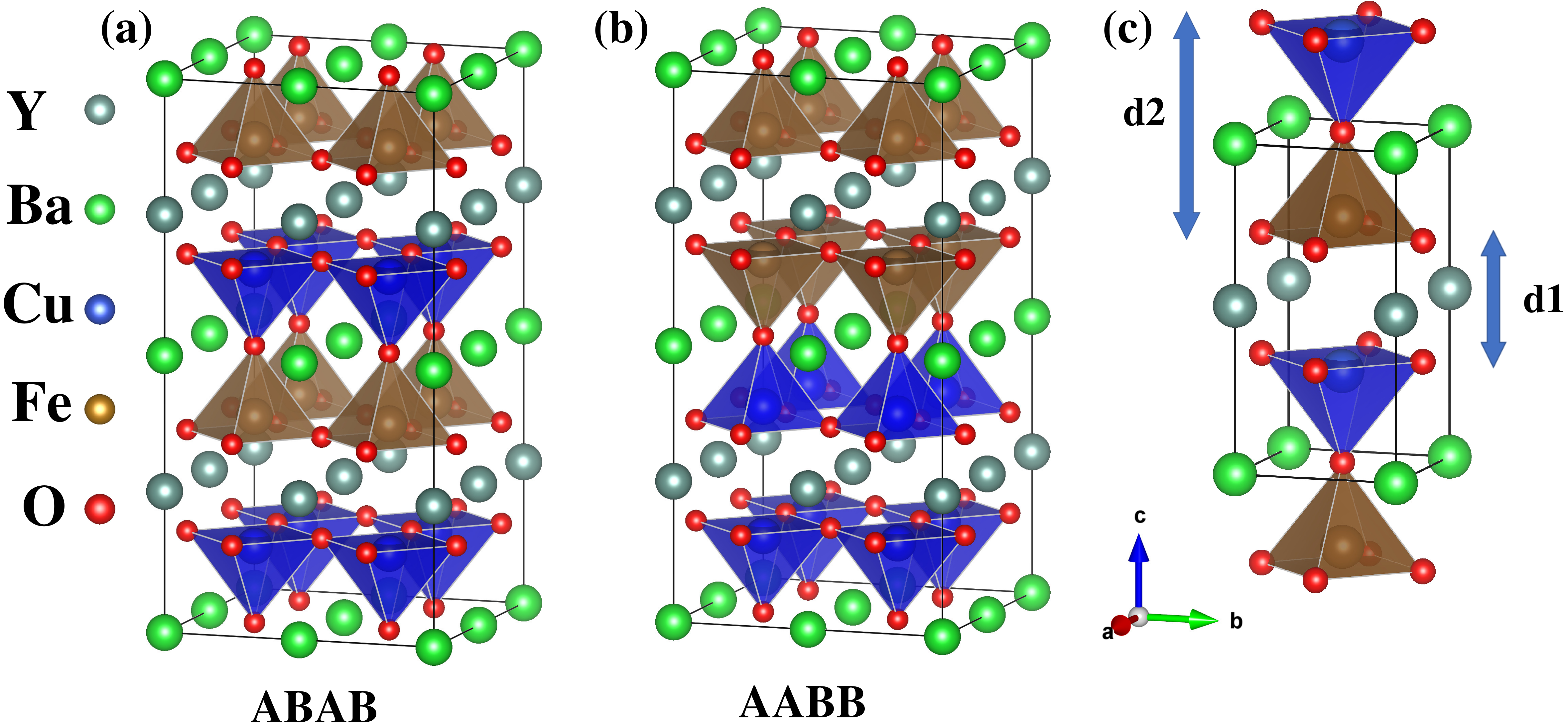}
    \caption{A $2\times2\times2$ YBCFO supercell is shown with (a) ABAB and (b) AABB Fe/Cu arrangements respectively. (c) One unit cell of YBCFO is shown where inter-bipyramidal distance (d$_1$) and bipyramidal thickness (d$_2$) are marked.}
    \label{fig1}
\end{figure}

Previous experimental reports\cite{Ruiz1998, Mombru1998, Caignaert1995, morin2015, Lai2017} suggest that YBCFO crystallizes in to two possible space groups P4/mmm and P4mm depending on Fe/Cu ion distribution where the former is centrosymmetric and the latter is non-centrosymmetric. In Fig. \ref{fig1} we present the crystal structure of YBCFO where bipyramidal layers of MO$_5$ (M = Fe, Cu) separated by Y$^{+3}$ layer can be seen. Due to very similar sizes of Fe$^{3+}$ and Cu$^{2+}$ ions both can go to the same site. However, it has been found that the experimentally observed ferromagnetic (FM) interaction within bipyramidal layers along $c$ direction is only possible if they are occupied preferentially by the Fe-Cu pairs\cite{morin2015}. Keeping this in view and considering experimentally observed antiferromagnetic (AFM) interaction in the $ab$-plane, we have considered two possible Fe/Cu arrangements as shown in Fig. \ref{fig1}: (a) ABAB arrangement where FeO$_5$-CuO$_5$ bipyramidal layers are stacked along $c$ direction in such a way that Fe and Cu layers alternate along $c$ and (b) AABB arrangement where FeO$_5$-CuO$_5$ bipyramidal layers are stacked along $c$ direction in such a way that two layers of Fe and two layers of Cu alternate along $c$. Fig. \ref{fig1}(c) depicts one unit cell of YBCFO, the bipyramidal thickness (d2) and inter-bipyramidal separation (d1).  

In a very recent experimental work, Zhang et al.\cite{zhang2021} have looked into the effect of doping of Mn at Fe sites, (i.e. YBaCuFe$_{1-x}$Mn$_{x}$O$_5$ (YBCFMO) with x=0, 0.01, 0.05, 0.1, 0.15 and 0.2) on the nature the spiral state in this system. The authors have observed that on increasing Mn doping the spiral plane tilts more towards $c$-axis from $ab$-plane even though the transition temperatures of spiral state as well as commensurate magnetic state (T$_{N2}$ and T$_{N1}$ respectively) decrease with doping. The tilting of spiral plane towards $c$-axis is a desirable property for having finite electric polarization in the system\cite{Katsura2005}. In this work we investigate the underlying mechanism behind such a change in spin moments' orientation through a detail electronic structure calculation using density functional theory.    

\section{Methods}
We used density functional theory (DFT), a first principles approach, to investigate the magnetic order, electronic structure, orbital state in Mn doped YBCFO. Vienna Abinitio Simulation Package (VASP) is used for DFT calculations which uses pseudopotentials and plane-wave basis sets\cite{vasp,vasp1}. We considered Perdew-Burke-Ernzerof generalized gradient approximation (PBE-GGA)\cite{gga} as exchange correlation functional for our calculations. We set the plane wave cut off to 600 \textit{eV} and sampled the Brillouin zone by $8\times8\times4$ Monkhorst-pack {\bf k}-grid points. YBCFO and YBCFMO are strongly correlated systems and so the effect of Coulomb correlation (U) for localized d-states cannot be ignored. We incorporated the on-site Coulomb correlation (U) and Hund's coupling strength (J) within the GGA+U approximation. For this purpose, we have set U (J) values at 5 (1), 5 (1) and 8 (0) eV for d-states of Fe, Mn and Cu respectively as used in previous literature\cite{morin2015}. We have considered YBCFO supercell as shown in Fig.\ref{fig1}(a-b) which is the parent compound where we substituted Fe by Mn to create the doped compounds YBCFMO with x = 0.125, 0.25. We used experimental\cite{zhang2021} lattice parameters for YBCFO (x = 0) and to find the lattice parameters for YBCFMO with x=0.125 and 0.25 we interpolated and extrapolated the experimental lattice parameters available for x=0.01, 0.05, 0.1, 0.15 and 0.2\cite{zhang2021}. For x=0.25 we have also performed c/a optimization. These lattice parameters for x=0.0, 0.125 and 0.25 are listed in Table \ref{table1}. We then performed full ionic relaxation until the Hellman-Feynman force reached 0.01 $eV/\AA{}$ for parent as well as doped compounds. Non-collinear calculations were performed within GGA+U+SO approximation where SO is the spin-orbit (SO)\cite{so} coupling. Magnetic exchange interactions are calculated using the method prescribed by Xiang et al.\cite{xiang}.

\section{Results and Discussion}

\subsection{Crystal Structure Analysis}
We have investigated in detail the effect of Mn doping on the YBCFO crystal structure using density functional theory calculations as discussed below. As it is observed experimentally that Mn ions have strong preference for Fe sites\cite{zhang2021}, we have performed calculations considering Mn ions substitute Fe rather than Cu in YBCFO. In Table \ref{table1} we list the lattice parameters and the total energies of ABAB and AABB arrangements (see Fig.\ref{fig1}) for undoped and Mn-doped cases considered in our calculations. Looking at Table \ref{table1} we observe that the lattice constants $a$ and $b$ decrease slowly but monotonically with Mn doping whereas the lattice constant $c$ increases with doping exactly similar to what is observed in the experiment\cite{zhang2021}. We further observe from Table \ref{table1} that ABAB arrangement has lower energy than AABB arrangement both in undoped and doped cases considered which is consistent with the P4mm symmetry observed in experiment. Therefore, in the discussion below we will mostly focus on ABAB structure. 
\begingroup
\setlength{\tabcolsep}{8pt} 
\renewcommand{\arraystretch}{1.5} 
\begin{table}[h!]
 \caption{Lattice parameters (in $\AA$) for $2\times2\times2$ supercell for YBCFO and YBCFMO are listed for different $x$ values.  $\Delta$E($x$) = E$_x$(AABB) - E$_x$(ABAB) is the energy difference, in $meV$, per formula unit between the two Fe/Cu ordering considered within GGA+U+SO approximation.}
 \vspace{2 mm}
\centering
 \begin{tabular}{c | c  c  c  c} 
\textit{x} & \textit{2a} & \textit{2b} & \textit{2c} & $\Delta$E($x$)\\
 \hline
 0.0 & 7.7493 & 7.7493 & 15.3253 & 72.01  \\
 0.125 & 7.7447 & 7.7447 & 15.3386 & 4.06 \\
 0.25 & 7.7405 & 7.7405 & 15.3503 & \\[1ex]
 \end{tabular}
       \label{table1}
\end{table}
\endgroup
\begin{figure}[ht!]
    \centering
    \includegraphics[width = 8.5 cm]{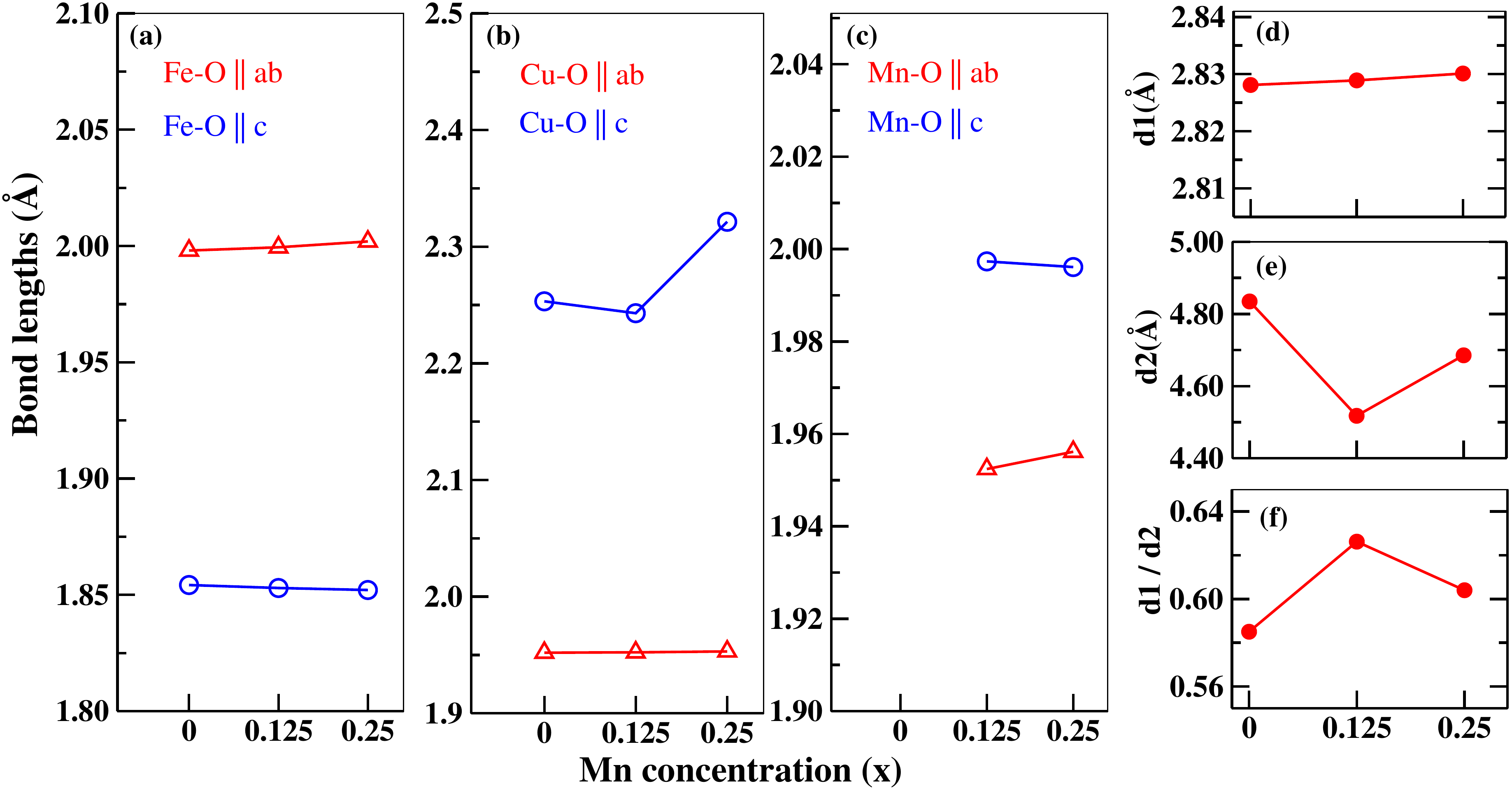}
    \caption{(a)-(c) show the calculated average basal and apical M-O bond lengths for M=Fe, Cu and Mn respectively as a function of doping concentration (x) in YBCFMO. (d), (e), and (f) represent the inter-bilayer separation (d$_1$), bilayer thickness (d$_2$), and their ratios (d$_1$/d$_2$), respectively as a function of doping. }
    \label{fig2}
\end{figure}

\begin{figure}[ht!]
    \centering
    \includegraphics[width = 8.5 cm]{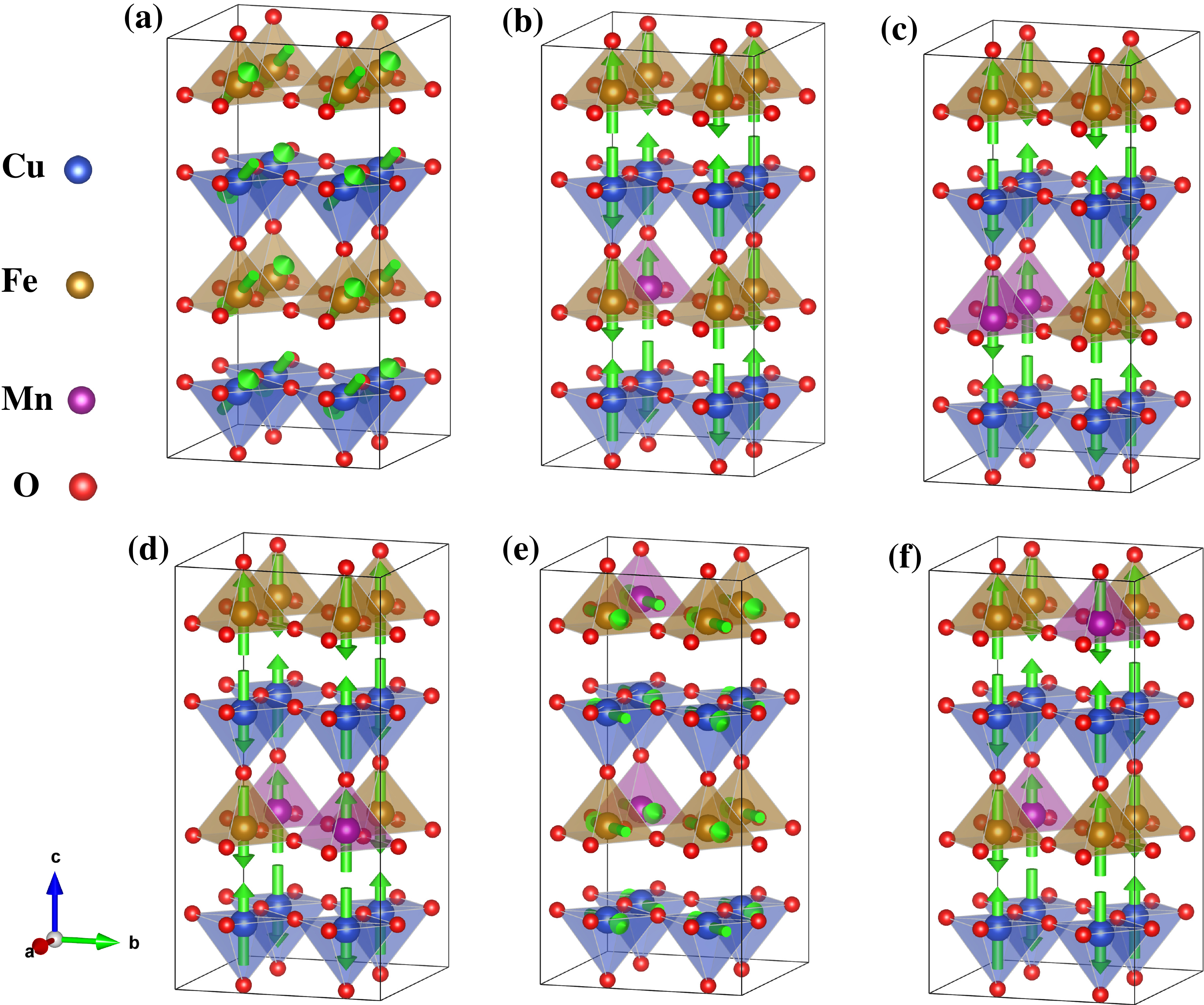}
    \caption{Ground state spin orientation in YBCFMO calculated within GGA+U+SO for (a) x=0, (b) x=0.125, (c) x=0.25 (confg-I), (d) x=0.25 (confg-II), (e) x=0.25 (confg-III) and (f) x=0.25 (confg-IV).}
    \label{fig3}
\end{figure}

To explore how local Jahn-Teller splitting is affected by doping we look at the average M-O bonds (M = Fe/Mn/Cu) in the $ab$-plane (basal) and along $c$-direction (apical). We present in Fig.\ref{fig2} (a)-(c) these bond lengths as a function of doping. We would like to note here that for x = 0.25 we have considered four different Mn distributions as shown in Fig. \ref{fig3}. The data used in Fig.\ref{fig2} for x = 0.25 are extracted from the lowest energy configuration out of the four. From Fig.\ref{fig2} we observe that in case of Fe-O, the basal bond is longer than the corresponding apical bond whereas it is the opposite in Cu-O case. This is consistent with the experimental observations by Zhang et al.\cite{zhang2021}. In case of Mn-O, the apical bond is longer than the basal ones. Thus the doping of Mn at Fe site would decrease the average basal M-O bond length and increase the average apical M-O bond length. The difference between apical and basal M-O bond lengths is maximum for Cu-O and minimum for Mn-O. Hence local Jahn-Teller effect is more in CuO$_5$ than in MnO$_5$. In Fig.\ref{fig2} (d)-(f) we present the variation of inter-bilayer separation (d$_1$) (see Fig.\ref{fig1}(c)) , bilayer thickness (d$_2$) and their ratio (d$_1$/d$_2$) respectively. Comparing with experiment\cite{zhang2021}, we find the behaviour of these three parameters are quite similar to that observe in experiment up to x = 0.125. However, their behaviour differs from experiment when we go from x = 0.125 to 0.25.

\subsection{Magnetic state and exchange interaction}
In the experimental measurements by Zhang et al.\cite{zhang2021}, it has been observed that Mn doping drives the system more towards a cycloidal spiral state where moments tend to move away from $ab$-plane unlike the helical spiral state of the parent compound observed previously by Lai et al.\cite{Lai2017}. To explore the effect of Mn doping on the spin orientation, we performed total energy calculations for various non-collinear spin orientations such as [100], [110],[001] and [111] within GGA+U+SO approximation considering the commensurate magnetic state. The total energy comparison is presented in Table \ref{table2}. Here for x = 0.25 case we have considered four different Mn distributions (see Fig. \ref{fig3}, denoted as confg-I, confg-II, confg-III and confg-IV).  

  \begingroup
\setlength{\tabcolsep}{6pt} 
\renewcommand{\arraystretch}{1.5} 
\begin{table}[h!]
 \caption{Total energy comparison among various orientations of spin moments for different Mn doping values calculated within GGA+U+SO approximation.}
\vspace{2 mm}
\centering
 \begin{tabular}{c c c c c}
 \multicolumn{1}{c}{}&\multicolumn{4}{c}{Orientation of spin moments}{}\\
Doping (x)   & [100] & [110] & [001] & [111]\\[1ex]
0.0             & 0.44  & 0.47  & 1.11  & \textbf{0.0}\\
0.125           & 1.84  & 1.94  & \textbf{0.0} & 1.42\\
0.25 (confg-I)       & 3.46  & 3.69  & \textbf{0.0}  & 2.59\\
0.25 (confg-III)      & \textbf{0.0} & 0.05 & 0.55 & 0.50\\[1ex]
\end{tabular}
\label{table2}
\end{table}
\endgroup

The ground state spin orientation is shown for various undoped and doped cases considered in our calculations in Fig. \ref{fig3}. We observe that in the parent compound the spins prefer to lie in the $ab$-plane if we consider AABB stacking as reported previously\cite{Dey2018}. However, if we consider ABAB stacking spins prefer to lie along [111] direction (see Fig \ref{fig3}(a)) which is consistent with data reported by Zhang et al.\cite{zhang2021}. With Mn doping of 12.5$\%$ (x = 0.125), the spin moments are seen to reorient towards $c$-axis (see Fig. \ref{fig3}(b)). For x = 0.25, the ground state spin orientations for four different Mn configurations (confg-I to confg-IV) are shown in Fig. \ref{fig3}(c)-(f) respectively. We observe that except confg-III (i.e. Fig. \ref{fig3}(e)) which is not the lowest energy state, all the remaining three configurations have their spin easy axis oriented towards $c$-axis. In confg-III the spin easy axis lies in the $ab$-plane.  Amongst these four configurations, confg-I (Fig. \ref{fig3}(c)) has the lowest energy. Therefore, we also conclude from our calculations that with Mn doping the spin easy axis is driven towards $c$ axis as observed in the experiment. To understand the spin reorientation with Mn doping, we evaluated various exchange interactions as listed in Table III using the method prescribed by Xiang et al.\cite{xiang}. 
 \begingroup
\setlength{\tabcolsep}{5pt} 
\renewcommand{\arraystretch}{1.5} 
\begin{table*}
\caption{Calculated nearest neighbour magnetic exchange interaction (J), in meV, within GGA+U.  Here J$_{ab}$, J$_{c}^{BP}$, and J$_{c}^{IP}$ denote the interaction between magnetic moments within ab-plane, along $c$ within bi-pyramids (BP), and along $c$ inter-bilayer (IP) respectively. Positive sign indicate antiferromagnetic interaction and negative sign indicate ferromagnetic interaction.}
 \vspace{2 mm}
\centering
 \begin{tabular}{c c c c c}
 \hline\hline
\textit{J} & \textit{x}$\xrightarrow{}$ \textit{0.0} & \textit{0.125 }& \textit{0.25 (confg-I)} & \textit{0.25 (confg-III)} \\
 \hline
$J_{ab}$&$J_{Cu-Cu}$ = 200.09 & $J_{Cu-Cu}$ = 183.08  & $J_{Cu-Cu}$ = 165.26    & $J_{Cu-Cu}$ = 151.65 \\
        &$J_{Fe-Fe}$ = 14.54  & $J_{Fe-Fe}$ = 12.92   & $J_{Fe-Fe}$ = 15.03     & $J_{Fe-Fe}$ = 7.19     \\
                            & & $J_{Fe-Mn}$ = 4.18    & $J_{Mn-Mn}$ = -0.15     & $J_{Fe-Mn}$ = 3.71   \\
	&		      &				& $J_{Mn-Fe}$ = 4.33    &			\\[1ex]
\hline
$J_{c}^{BP}$& $J_{Fe-Cu}$ = -1.79 & $J_{Fe-Cu}$ = -1.94 & $J_{Fe-Cu}$ = -2.11  & $J_{Fe-Cu}$ = 5.39     \\
                                & & $J_{Cu-Mn}$ = -1.17 & $J_{Cu-Mn}$ = -0.89  & $J_{Cu-Mn}$ = 2.68     \\[1ex]
\hline
$J_{c}^{IP}$&$J_{Fe-Cu}$ = 1.29   & $J_{Fe-Cu}$ = 1.18  & $J_{Fe-Cu}$ = 1.16   & $J_{Fe-Cu}$ = -8.31    \\
                            &     & $J_{Cu-Mn}$ = -2.42 & $J_{Cu-Mn}$ = -2.91  & $J_{Cu-Mn}$ = -0.77   \\
                                                                         & & & &			\\[1ex]
 \hline
 \hline
\end{tabular}
\label{table3}
\end{table*}
\endgroup
\begin{figure}[ht!]
    \vspace{0.2 cm}
    \centering
    \includegraphics[width = 8.5 cm]{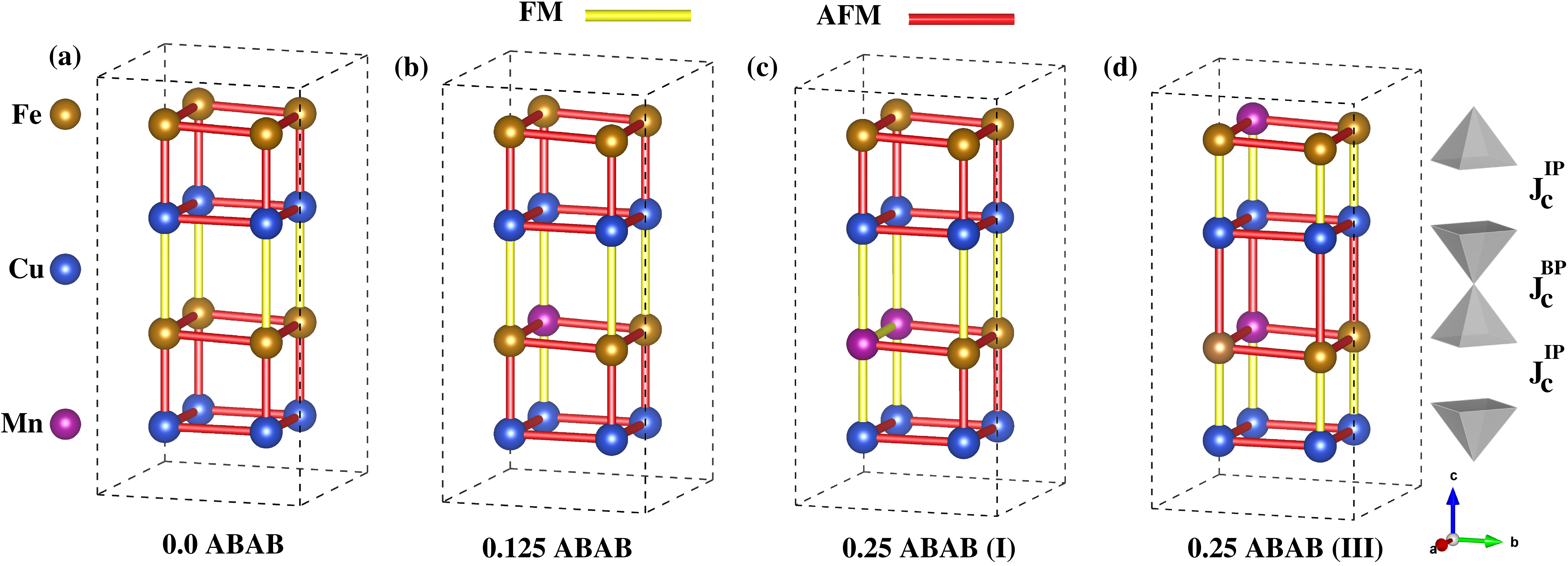}
    \caption{Depicting the nature of interaction (by colored bonds) between the nearest neighbour magnetic moments (on the basis of Table III) for the for (a) x=0, (b) x=0.125, (c) x=0.25 (confg-I), (d) x=0.25 (confg-III). Yellow and red colored bonds represent the ferromagnetic (FM) and anti-ferromagnetic (AFM) interaction, respectively. Stacking of pyramids along $c$ is shown on the right side for reference.}
    \label{fig4}
\end{figure}

From Table \ref{table3}, we observe that the $ab$-plane exchange interactions between Fe-Fe, Cu-Cu and Fe-Mn remain antiferromagnetic in nature with Mn doping similar to the parent compound. The largest interaction among all is Cu-Cu exchange which is seen to decrease with Mn doping. This is in line with the experimental observation that with Mn doping the magnetic transition temperatures of both commensurate and incommensurate phases decrease\cite{zhang2021}. Looking at the exchange interactions along $c$ direction we find that for parent compound there exists ferromagnetic interaction within the bi-pyramid (J$_c$$^{BP}$, see Fig.\ref{fig4}) whereas inter-bilayer (J$_c$$^{IP}$) it is antiferro which is consistent with the experimentally observed commensurate magnetic structure\cite{morin2015}. When we dope 12.5$\%$ Mn at Fe sites (x=0.125), we observe that inter-bilayer Cu-Mn exchange becomes ferromagnetic whereas the corresponding Fe-Cu exchange remains antiferro. The magnitude of this inter-bilayer Cu-Mn exchange is also larger than other Fe/Mn-Cu exchange along $c$-direction. Therefore, this causes frustration in the commensurate magnetic interactions forcing the moments to cant away. When we move to higher Mn doping (i.e. x = 0.25, confg-I) we observe similar behaviour as in x = 0.125. Interestingly, in confg-III of x = 0.25, we observe that along c direction the exchange interactions within the bi-pyramid have become antiferromagnetic whereas the inter-bilayer ones ferromagnetic; just opposite to the case in parent compound. These observations have been depicted in Fig.\ref{fig4}. In this case the frustration is therefore released. This can be further understood from the electronic structure which we present in the following section.

\subsection{Electronic structure}                

\begin{figure}[ht!]
    \centering
    \includegraphics[width = 8.5 cm]{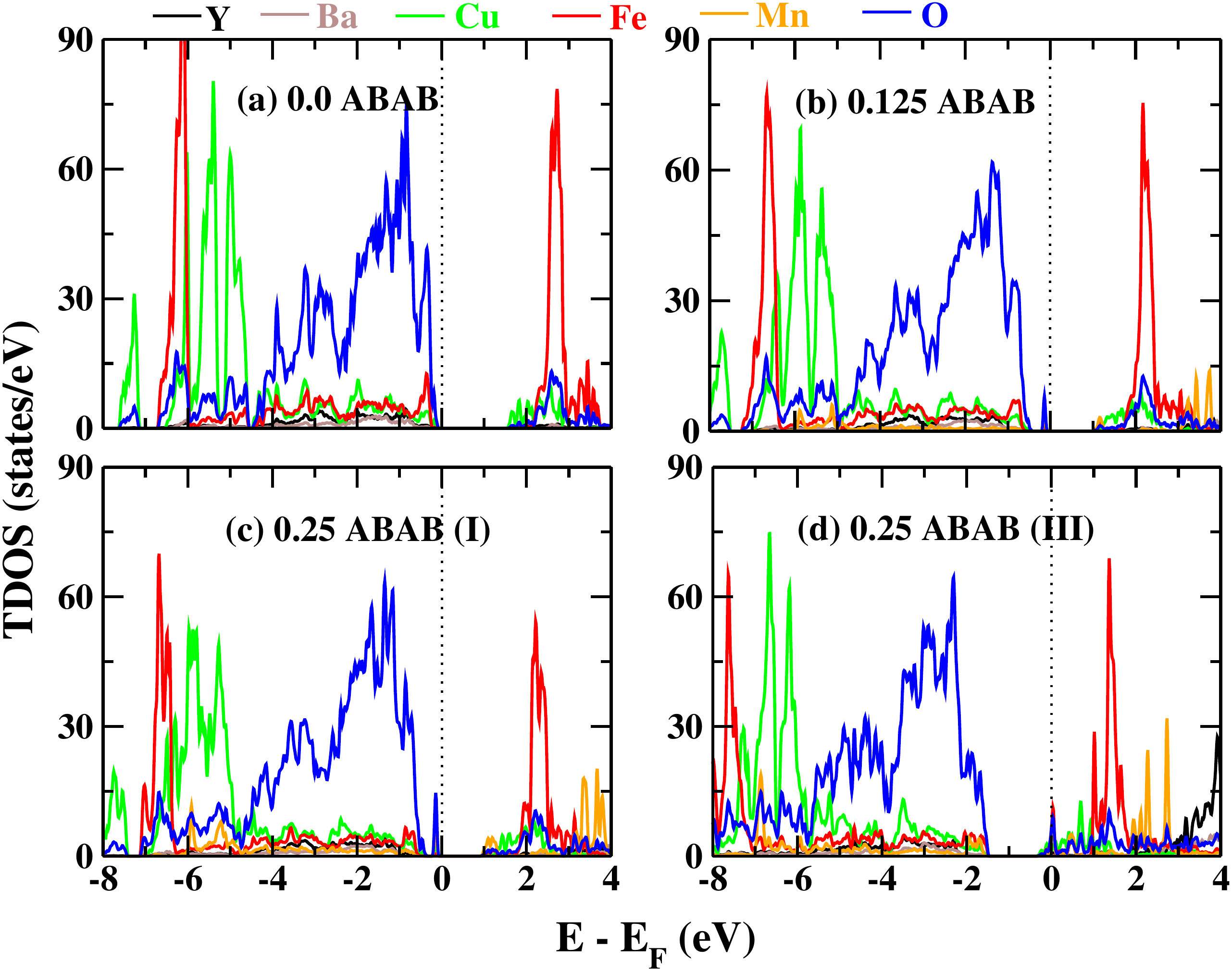}
    \caption{Atom projected total density of states (TDOS) calculated within GGA+U+SO for (a) x=0, (b) x=0.125, (c) x=0.25 (confg-I) and (d) x=0.25 (confg-III).}
    \label{fig5}
\end{figure}

We performed electronic structure calculations for undoped and Mn doped YBCFO for their respective magnetic ground state within GGA+U+SO approximation as shown in Fig.\ref{fig3}. We present in Fig.\ref{fig5} the total density of states (TDOS) where we observe insulating state for the undoped (x = 0.0) and 12.5$\%$ Mn doped (x = 0.125) case (see Fig.\ref{fig5}(a) and (b)) whereas for 25 $\%$ Mn doped case (x=0.25) out of the four configurations we have studied, three (confg-I,II and IV) are insulating and one (confg-III, Fig.\ref{fig3}(e)) is metallic. In Fig.\ref{fig5}(c) we present the TDOS for the lowest energy configuration (i.e. confg-I, \ref{fig3}(c)) at x=0.25 and in Fig.\ref{fig5}(d) we present the same for highest energy configuration (i.e. confg-III). We further observe that around the Fermi level Cu/Fe/Mn states are present which strongly hybridized with oxygen states. To understand the mechanism of different ground state spin orientations as discussed in the previous section and also different transport behaviour (insulating/metallic) we looked into the details of electronic structure around the Fermi level.
\begin{figure}[ht!]
    \centering
    \includegraphics[width = 8.5 cm]{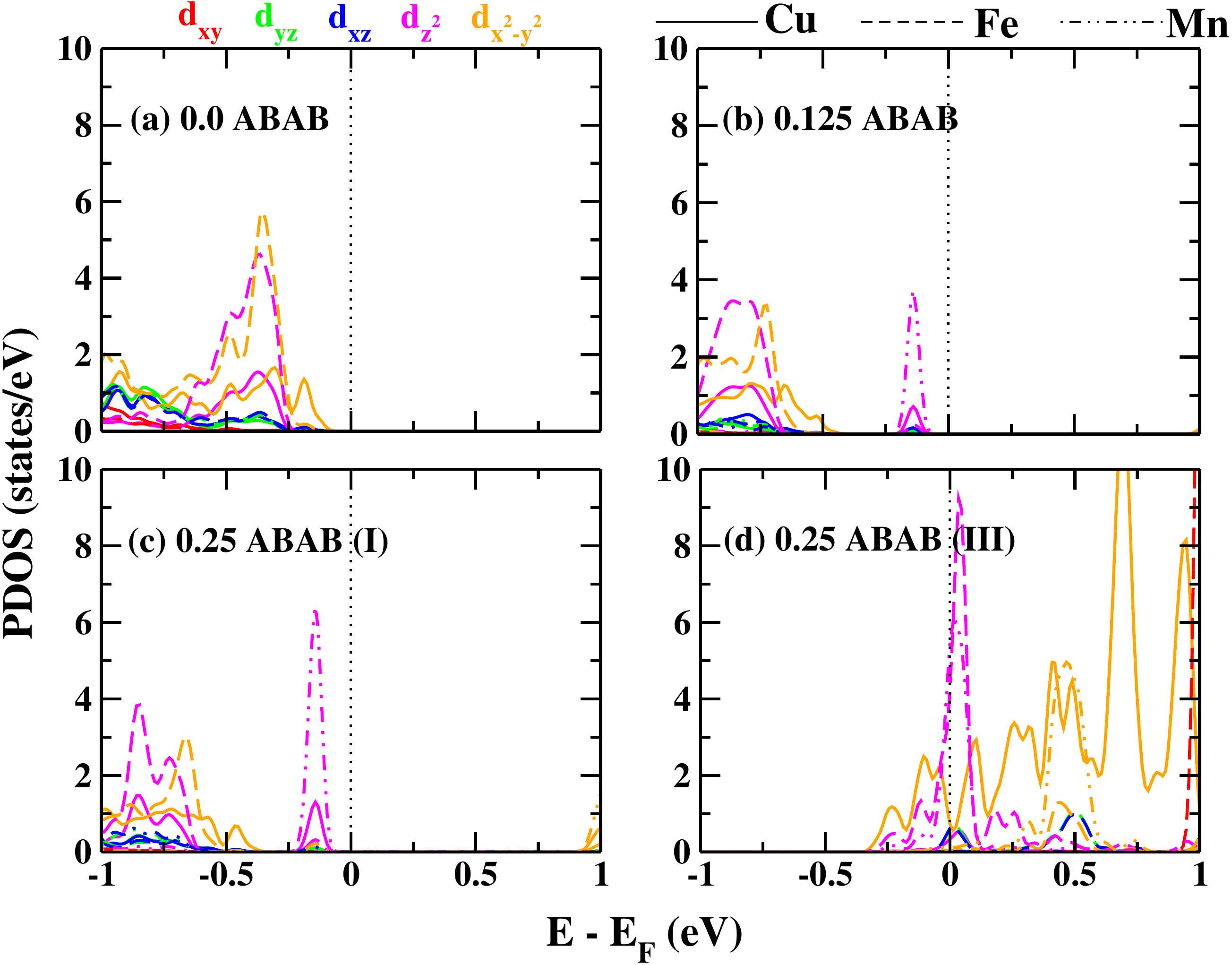}
    \caption{Orbital projected partial density of states (PDOS) calculated within GGA+U+SO for (a) x=0, (b) x=0.125, (c) x=0.25 (confg-I) and (d) x=0.25 (confg-II).}
    \label{fig6}
\end{figure}
\begin{figure}[ht!]
    \centering
    \includegraphics[width = 8.5 cm]{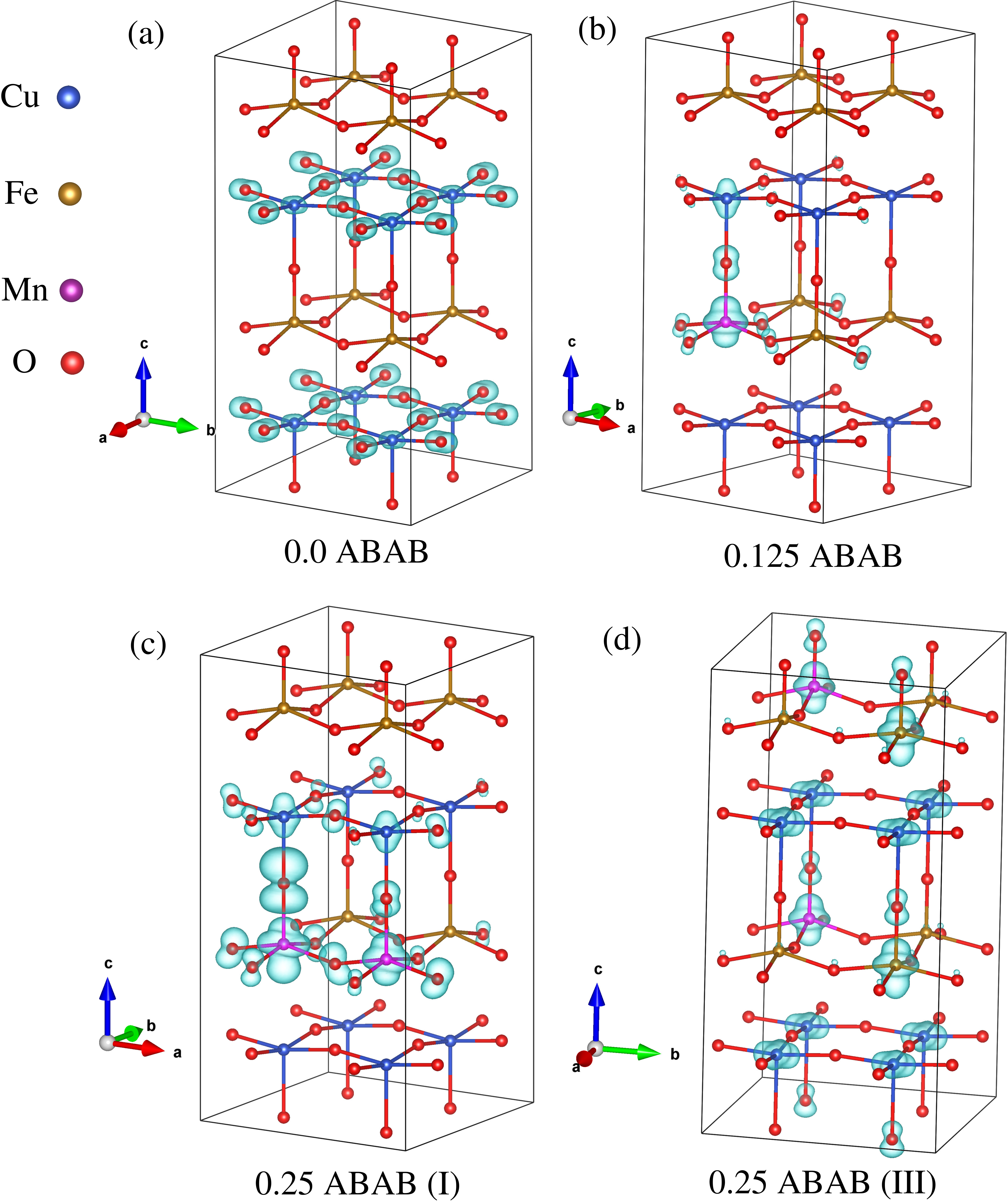}
    \caption{3D electron density plotted for the energy range : -0.25 eV to Fermi level (0 eV) showing the occupied orbitals at various sites for (a) x=0, (b) x=0.125, (c) x=0.25 (confg-I) and (d) x=0.25 (confg-III).}
    \label{fig7}
\end{figure}
We calculated the orbital projected partial density of states (PDOS) as well as 3D electron densities which we present in Fig.\ref{fig6} and Fig.\ref{fig7}, respectively. As we see in Fig.\ref{fig6}(a) the Cu d$_{x^2-y^2}$ is the highest occupied orbital in the parent compound (x = 0.0). Upon doping Mn at x = 0.125 we observe that Mn d$_{z^2}$ becomes the highest occupied orbital which hybridizes with Cu d$_{z^2}$ and oxygen p-orbital. Same situation is also observed at x = 0.25 with confg-I with even stronger hybridization between Mn d$_{z^2}$, Cu d$_{z^2}$ and Oxygen p$_z$ orbital. This can be clearly seen in calculated 3D electron densities in the energy range -0.25eV to Fermi level (0 eV) which are shown in Fig.\ref{fig7}(a) to (c) for x = 0.0, 0.125 and 0.25 respectively. Interestingly, in confg-III at x=0.25 (see Fig.\ref{fig6}(d)) we observe that there is a charge transfer from the occupied Mn d$_{z^2}$ orbital to unoccupied Fe d$_{z^2}$ and Cu d$_{x^2-y^2}$ orbitals making the system metallic. 
From Fig.\ref{fig7}(d) we can clearly see the occupied Cu d$_{x^2-y^2}$ orbitals similar to the parent compound. Thus we can conclude that with Mn doping Cu d$_{z^2}$ becomes the highest occupied orbital due to its hybridization with Mn d$_{z^2}$ in contrast to the parent compound where Cu d$_{x^2-y^2}$ orbital was the highest occupied orbital. The occupancy of out of plane oriented Cu d$_{z^2}$ orbital in the doped compounds in place of the planar Cu d$_{x^2-y^2}$ of parent compound facilitates the spin moments canting towards $c$ direction upon doping Mn as observed in the experiment. In confg-III of x = 0.25 as the Cu d$_{x^2-y^2}$ gets occupied through charge transfer from Mn, the spin moments prefer to lie in the $ab$-plane as observed in our calculation (see Fig.\ref{fig3}(e)). Therefore, the orbitals play a major role in driving the spin moments out of plane in YBCFMO.

\section{Conclusions}

In summary, we carried out a thorough investigation to understand the mechanism behind spin reorientation upon Mn doping at Fe sites in YBCFO using density functional theory calculations. In this context, we studied the non-collinear magnetic order, electronic structure and orbital state of doped and undoped YBCFO. Through a detailed analysis of optimized crystal structures of the undoped (x=0.0) and doped (x=0.125 and 0.25) compounds, we have found the local Jahn-Teller distortions of CuO$_5$, FeO$_5$ and MnO$_5$ pyramids which are consistent with experimental results observed by Zhang et al.\cite{zhang2021}. From our total energy calculations we have obtained the lowest energy configurations for Fe/Cu distribution as well as Mn distribution in the three compounds. Considering the respective lowest energy configurations we have performed non-collinear magnetic calculations within GGA+U+SO approximation to obtain the ground state magnetic order. Considering the commensurate magnetic structure for all the three compounds, we observe that the spin moments indeed prefer to align along crystallographic $c$ direction in the doped compounds whereas in parent compound they are aligned along (111) direction. This is consistent with the trend observed in the experimental measurements. Further, our estimates of nearest neighbour magnetic exchange interactions in the $ab$-plane and along $c$ direction reveal that the Cu-Cu exchange in $ab$-plane which is AFM and the largest amongst all, decreases with Mn doping. This is again in line with the experimental observation of lowering of transition temperature upon doping. More interestingly, we find that Mn doping gives rise to a frustrating inter-bilayer ferromagnetic exchange which otherwise should have been antiferromagnetic in the commensurate magnetic structure. This we believe may be one of the reasons behind spins canting away from their corresponding orientation in the parent compound. The other possible reason is the change in the nature of highest occupied Cu orbital. In parent compound, the orbital is d$_{x^2-y^2}$ which is a planar orbital but in the doped compound due to hybridization with Mn d$_{z^2}$, Cu d$_{z^2}$ becomes the highest occupied orbital. This change in orbital state we believe drives the spin moments out of plane towards $c$ direction in the doped compounds. We found one particular configuration of Mn distribution (not lowest energy) in x=0.25 case which is metallic due to charge transfer from Mn d$_{z^2}$ to unoccupied Cu d$_{x^2-y^2}$ in the conduction band. Interestingly, we find that in this case also the spin moments prefer to lie in the $ab$-plane. Therefore, the spin orientation in the doped compound appears to be orbitally driven aided by the frustrating exchange interactions.

\section{Acknowledgement}
MS would like to acknowledge the Ministry of Education, India for the research fellowship. MS and TM acknowledge the National Supercomputing Mission (NSM) for providing computing resources of ‘PARAM Ganga’ at the Indian Institute of Technology Roorkee, which is implemented by C-DAC and supported by the Ministry of Electronics and Information Technology (MeitY) and Department of Science and Technology (DST), Government of India. TM acknowledeges Science and Engineering Research Board (SERB), India for funding support through MATRICS research grant (MTR/2020/000419).

\end{document}